\documentclass[review]{elsarticle}

\usepackage{mathrsfs}
\usepackage{graphicx}
\usepackage{amsmath}
\usepackage{amsfonts}
\usepackage{amssymb}
\usepackage{amsthm}
\usepackage{textcomp}
\usepackage{color}
\usepackage{geometry}
\usepackage[normalem]{ulem}
\usepackage{placeins}
\usepackage{bm}
\usepackage{wasysym}
\usepackage{titlesec}
\usepackage{txfonts}
\usepackage{graphicx}
\usepackage{subfigure}
\usepackage{braket}
\usepackage{verbatim}
\usepackage{lineno,hyperref}

\newcommand{\<}{\langle}
\renewcommand{\>}{\rangle}
\newcommand\dd{~{\rm d}}

\newcommand\R{\mathbb{R}}

\begin{document}
\nolinenumbers

\begin{frontmatter}

\title{An extended plane wave framework for the electronic structure calculations of twisted bilayer material systems}

\author{Xiaoying Dai}
\address
{LSEC, Institute of Computational Mathematics and Scientific/Engineering Computing, Academy of Mathematics and Systems Science, Chinese Academy of Sciences, Beijing 100190, China}
\address
{School of Mathematical Sciences, University of Chinese Academy of Sciences, Beijing 100049, China}
\ead{daixy@lsec.cc.ac.cn}

\author{Aihui Zhou}
\address
{LSEC, Institute of Computational Mathematics and Scientific/Engineering Computing, Academy of Mathematics and Systems Science, Chinese Academy of Sciences, Beijing 100190, China}
\address
{School of Mathematical Sciences, University of Chinese Academy of Sciences, Beijing 100049, China}
\ead{azhou@lsec.cc.ac.cn}

\author{Yuzhi Zhou\corref{correspondingauthor}}
\address
{CAEP Software Center for High Performance Numerical Simulation, Beijing 100088, China}
\address
{Institute of Applied Physics and Computational Mathematics, Beijing 100088, China}

\cortext[correspondingauthor]{Corresponding author}
\ead{zhou\_yuzhi@iapcm.ac.cn}

\date{\today}

\begin{abstract}
 In this paper, we propose an extended plane wave framework to make the electronic structure calculations of the twisted bilayer 2D material systems practically feasible. Based on the foundation in [Y. Zhou, H. Chen, A. Zhou, J. Comput. Phys. 384, 99 (2019)], following extensions take place: (1) an tensor-producted basis set, which adopts PWs in the incommensurate dimensions, and localized basis in the interlayer dimension, (2) a practical application of a novel cutoff techniques we have recently developed, and (3) a quasi-band structure picture under the small twisted angles and weak interlayer coupling limits. With (1) and (2) now the dimensions of Hamiltonian matrix are reduced by about 2 orders of magnitude compared with the original framework. And (3) enables us to better organize the calculations and understand the results. For numerical examples, we study the electronic structures of the linear bilayer graphene lattice system with the magic twisted angle ($\sim 1.05^{\circ}$). The famous flat bands have been reproduced with their features in quantitative agreement with those from experiments and other theoretical calculations. Moreover, the extended framework has much less computational cost compared to the commensurate cell approximations, and is more extendable compared to the traditional model hamiltonians and tight binding models. Finally this framework can readily accommodate nonlinear models thus will laid the foundations for more effective yet accurate Density Functional Theory (DFT) calculations.
\end{abstract}


\end{frontmatter}


\section{Introduction}
\label{sec:introduction}

The lack of periodicity poses a major difficulty for the theoretical studies of the electronic properties of twisted 2D material systems, where an abundance of fascinating correlation phenomena and physics are hosted \cite{Cao18,Cao18sc,Kennes2021,Andrei2020}. A common approach is approximating them by commensurate supercells, such that the periodic boundary condition is restored and the Bloch theorem can be applied. However, in many cases the size of an acceptable commensurate cell is overly large and generally out of the capability of most computational source. A good example is the magic angle ($\sim 1.05^{\circ}$) twisted bilayer graphene (TBG). Yet the smallest supercell to approximate this TBG system requires more than 10,000 atoms, which puts serious challenge for DFT calculations using PW basis \cite{Carr2020}. Meanwhile, this approximation also prohibits a more systematic error analysis.

Alternatively one could switch to tight binding (TB) basis, which relieves the computational cost to some extent.  We further note that recently there have emerged improved TB models and corresponding efficient numerical algorithms, which treat the incommensurate system from rigorous mathematical foundations \cite{massatt17,massatt18,Massatt2020}. Even though, one still needs to worry about parameters in the hamiltonian, constructing localized orbitals, and convergence issues. In many scenarios one would like to have similar footings for the PW basis, which is in nature almost free from previous issues, to better facilitate the theoretical studies of incommensurate material systems.

Our previous work has partially fulfilled that goal by introducing a general PW framework for the quantum eigenvalue problem of the incommensurate systems, where major theoretical aspects have been discussed \cite{zhou19}. However, the number of PWs needed to discretize the Hamiltonian is huge, due to the fact that it is in principle a higher dimensional problem. Effective schemes to reduce the dimensions of the incommensurate problems are therefore needed. With that in mind, layer-splitting methods have been developed to simulate the time evolution of quantum states in the incommensurate potentials, where the states are evolving sequentially by each composed periodic potentials rather than the full incommensurate one. This reduces the original problem to several low dimensional sub-problems \cite{Wang2021}.

On the other hand, when we were trying to understand the extended-localized transitions in the incommensurate systems, we observed that the low-lying eigenstates in the higher dimensional reciprocal space were distributed in a very directional manner \cite{ChenLoc2021}. This gave us hints of better cutoff scheme. Later on, we solidified this idea by rigorous numerical analysis and efficient algorithms \cite{wangtCutoff}. By splitting the cutoffs into a traverse one and a quadratically increasing one, we showed in the numerical experiments that a much smaller PWs basis set as well as faster convergence can be achieved. Yet this scheme has not been applied to more practical calculations in the 2D material systems.

Besides that, previous work commonly neglect the interlayer dimension (denoted as $\emph{z}$) for simplicity \cite{zhou19,wangtCutoff,Wang2021}. But in the TGB, the interlayer coupling is crucial to give rise to the flat bands, and consequently the correlated phenomena such as the superconductivity and Mott insulator \cite{Cao18,Cao18sc}. However when trying to put back this dimension under the full PW basis, one needs a very long $z$ cell to separate the spurious image interactions from the periodic boundary condition. As a result, this leads to very large number of PWs from $z$ dimension. And the problem is further amplified by the incommensurate nature since we already have a higher dimensional problem in the $x$, $y$ directions. A natural way to improve is to replace the PW by localized functions in $z$ direction, though many formula in the framework have to be adjusted accordingly.


Based on the above discussions, in this paper we extend our PW framework in the following aspects: (1) a tensor-producted basis with PWs in the incommensurate dimensions and localized basis in $\emph{z}$ direction, with which we reformulated the eigenvalue problem; (2) the application of our newly developed sampling and cutoff techniques in more practical systems; (3) a quasi-band structure formulated under the small twisted angles and weak interlayer coupling limits. In principle, with (1) and (2) we can treat the full electronic structure calculations of twisted 2D material systems with moderate computational cost affordable by most modern computers. (3) is also equivalent to the mini Brillouin zone (BZ) description used by most continuum model Hamiltonian calculations \cite{Carr2020,MacD2011}. And even though ergodicity is more fundamental, with (3) we can better organize our calculations as well as understand the results. Also it enables us to compare with those band structure results from commensurate cell approximations or model hamiltonians.

As for numerical examples, the electronic structures of the linear TBG system with twist angle $\sim 1.05^{\circ}$ are studied. We have reproduced the famous flat bands with key features in quantitative agreement with those from experiments and other theoretical models \cite{Andrei2020,Carr2020,Li2010,massat2021}. Moreover, our calculations have much less computational cost compared to the commensurate cell approximations. Also it is more extendable compared to the traditional model hamiltonians and tight binding calculations, since in this work we utilize at most two parameters, and in principle it can be made fully \emph{ab initio}. Finally, this framework can readily accommodate nonlinear terms like Hartree energy and exchange-correlation energy and will laid the foundations for more effective and accurate DFT calculations for the incommensurate 2D material systems, which will be addressed in our future work.

This paper is organized as follow: in Sec. 2, the incommensurate quantum eigenvalue problem is reformulated using the PW$\otimes$localized-function basis set. In Sec. 3, the practical application of new cutoff techniques are discussed. In Sec. 4, we discuss the quasi-band structure that can be formulated under the small twisted angles and weak interlayer coupling limits. In Sec. 5, the linear TGB system with magic twisted angle is taken as numerical examples, and we show and discuss the results on the electronic structures and density distributions. In the last section, conclusions are given.

\section{Problem under tensor-producted basis}
\label{sec:basis}
The linear eigenvalue problem of a twisted bilayer system can be written as:
\begin{equation}
\label{eigen}
\left( -\frac{1}{2}\Delta + V_1(\textbf{x},z) + V_2(\textbf{x},z) \right)~\Psi(\textbf{x},z) = E~\Psi(\textbf{x},z) ,
\end{equation}
where the interlayer dimension $\emph{z}$ has been explicitly included. $\textbf{x}$ compactly represents the inplane dimensions $x$ and $y$. By definition, $V_1(\textbf{x},z)$ and $V_2(\textbf{x},z)$ are periodic potentials in $\textbf{x}$:
$V_j(\textbf{x} + n \textbf{R}_j,z)=V_j(\textbf{x},z)$ with $\textbf{R}_j$ the lattice constants for layer $j=1,2$. And incommensurate constraint is given by:
\begin{equation}
\textbf{R}_1\cup\textbf{R}_2 + \tau = \textbf{R}_1\cup\textbf{R}_2 \quad \Leftrightarrow \quad \tau=\pmb{0}\in\R^d .
\end{equation}

Under the full PW basis, a common practice to model slab systems (layers, surfaces and interfaces) is adding a vacuum layer (usually about 20 \AA) in $z$ direction, which serves to screen the spurious interactions between image parts. This will make the number of PWs a few orders of magnitude larger compared to bulk calculations and may cause some convergence issues \cite{Zhou18Kerker}. These problems are further amplified in the incommensurate systems since one already has a large number of PWs in $\textbf{x}$. Here, we adopt a tensor-product basis set with PWs in $\textbf{x}$ and some localized functions in $z$:
\begin{equation}
\Omega_{in} = \frac{1}{\sqrt{A}} e^{i\textbf{k}_i \cdot \textbf{x} } \xi_n(z) ,
\end{equation}
where $A$ is some large area within the $\textbf{x}$ dimensions in which all coupled PWs are normalized. With properly chosen localized functions, such as atomic-like orbitals or finite elements, a much small basis set as well as a comparable accuracy with PW code can be expected.

Now we reformulate the eigenvalue problem under the new basis. For the PW part, elements in the set $\{\textbf{k}_i\}$ are still differed by the sum of two sets reciprocal lattice vectors $\{\textbf{G}_1, \textbf{G}_2\}$ and $\{\textbf{Q}_1, \textbf{Q}_2\}$, as derived in \cite{zhou19}:
\begin{equation}
\label{pw}
\textbf{k}_i - \textbf{k}_j = n \cdot \textbf{G}_{1} + m \cdot \textbf{G}_{2} + k \cdot \textbf{Q}_{1} + l\cdot\textbf{Q}_{2} ,
\end{equation}
where $n, m, k, l$ are integers and $i, j$ should be viewed as compact notation of these integers. For simplicity of the presentation, we assume that the localized functions are orthogonal (though it is not necessary):
\begin{equation*}
\int^{\infty}_{-\infty} \xi^*_m(z) \xi_n(z) \dd z = \delta_{mn} .
\end{equation*}
And one can easily verify that $\{\Omega_{in}\}$ are also orthogonal.
The discretization of the Hamiltonian in \eqref{eigen} follows similar procedures in \cite{zhou19}, with some complexity comes in due to the localized functions. The matrix elements of kinetic operator take the form:
\begin{equation}
\label{kin}
\<\Omega_{jm}|\hat{T}|\Omega_{in}\> =  \delta_{ij} \bigg( \frac{1}{2}|\textbf{k}_i|^2+ \int^{\infty}_{-\infty} \xi^*_m(z) \frac{\dd^2}{\dd z^2} \xi_n(z) \dd z \bigg) .
\end{equation}
The matrix elements of each potential components can be calculated as:
\begin{eqnarray}
\label{preal}
\<\Omega_{jm}|V_l|\Omega_{in}\> &=& \frac{\delta_{\textbf{G}_l,\textbf{k}_{ij}}}{A_c}S_l(\textbf{k}_{ij}) \int^{\infty}_{-\infty} V^{at}_l(\textbf{k}_{ij},z) \xi^*_m(z) \xi_n(z) \dd z \\
\label{prec}
&=& \frac{\delta_{\textbf{G}_l,\textbf{k}_{ij}}}{A_c}S_l(\textbf{k}_{ij}) \int^{\infty}_{-\infty} V^{at}_l(\textbf{k}_{ij},k_z)  (\xi^*_m \cdot \xi_n)(k_z) \frac{\dd k_z}{2\pi},
\end{eqnarray}
where in \eqref{prec}:
\begin{equation*}
(\xi^*_m \cdot \xi_n)(k_z) = \int \xi^*_m(z) \xi_n(z) \textbf{e}^{-i k_z z} \dd z .
\end{equation*}
In the above equations, $\textbf{k}_{ij}$ is a compact notation for $\textbf{k}_i - \textbf{k}_j$, $A_c$ the unit cell area, $V^{at}_l$ the atomic potential and $S_l$ the corresponding structural factor. The integration of $z$ can either be performed directly in the real space \eqref{preal} or in the reciprocal space \eqref{prec}. The matrix elements of other ingredients in DFT, including the Hartree term, exchange-correlation term and nonlocal part of the potential can be derived similarly. Since we focus on the linear system in this paper, they will be reserved for our future studies.

With Eqs.~\eqref{kin}, \eqref{preal} and \eqref{prec}, one can in principle construct the full Hamiltonian matrix and solve for the eigenpairs. The above efforts relieve a large part yet unnecessary computational cost from the inefficient PW representation of the localized wavefunctions in $z$ direction. To further save the computational cost, we will briefly introduce our recently developed cutoff scheme and apply it to practical calculations. Some insights on the $k$-point sampling will be given as well, which helps to form the quasi-band structure description later on.

\section{k-point sampling and cutoff schemes}
As termed and discussed in \cite{zhou19,ChenLoc2021}, the coupled PWs $\{\textbf{k}_i\}$ in \eqref{pw} are ergodic, meaning that they densely and uniformly distributed in 2d reciprocal space for infinitely large enough cutoffs. Though one could argue that in principle single point sampling with large cutoff is enough, in many practical cases the multi $k$-point sampling is still preferred for faster convergence. This is best manifested in systems that are very close to periodic systems, for instance the small twisted angle bilayer systems to be discussed. First we briefly recap some basis on $k$-point sampling and ergodicity.

\begin{figure}[htb!]
\centering
	\includegraphics[width=15.0cm]{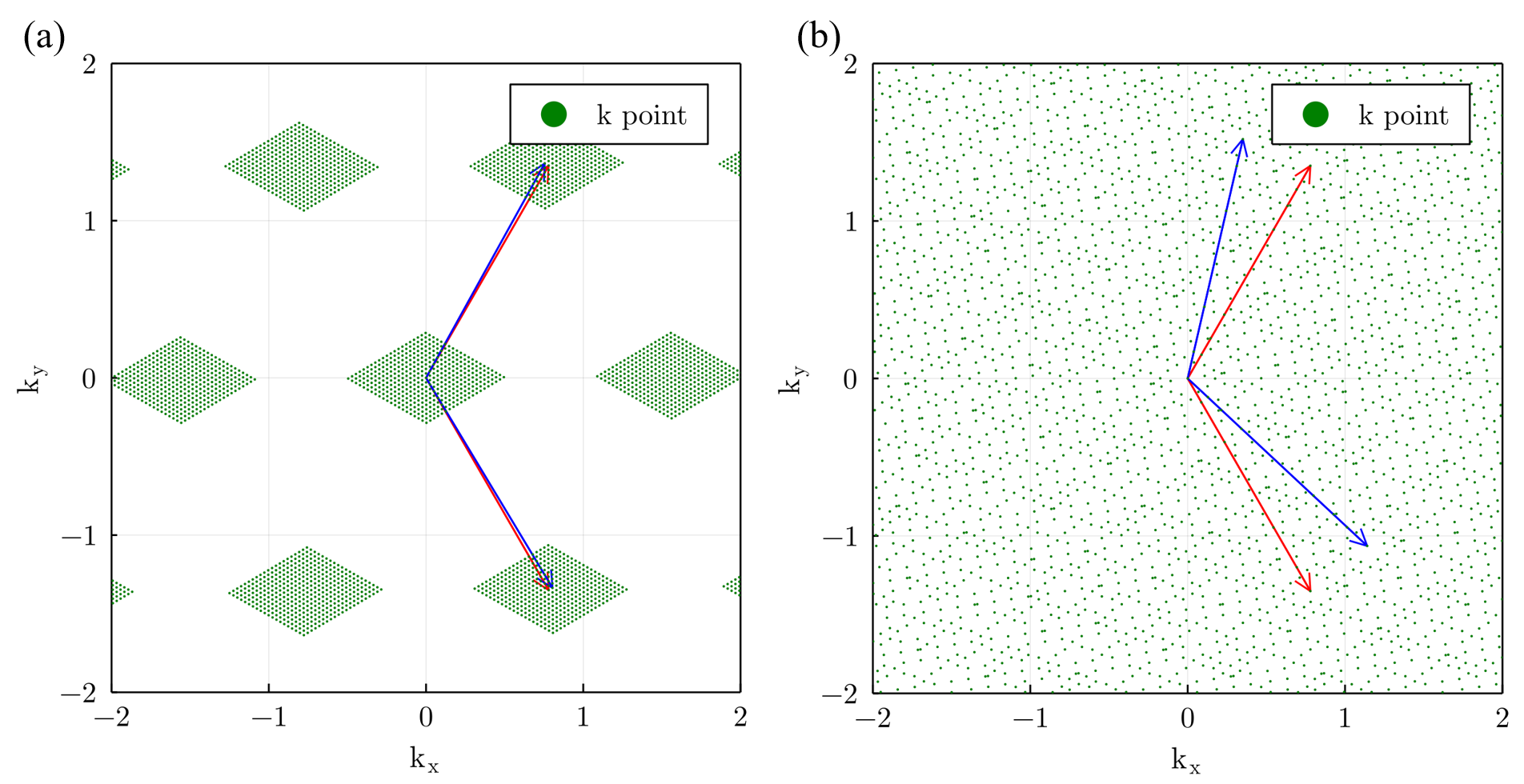}
	\caption{PW distribution in bilayer hexagonal systems with (a) $1.05^\circ$ and (b) $17^\circ$ twisted angles. The arrows indicate the original and twisted lattice vectors. The cutoff is $k_c = 6$. The scale is the atomic unit.}
	\label{fig:distr}
\end{figure}

In Fig.~\ref{fig:distr} we compare the distribution of PWs in bilayer hexagonal lattice systems with small ($1.05^\circ$) and large ($17^\circ$) twisted angles. As a variant of Eq.~\eqref{pw}, the PW sets $\{\textbf{k}^{i}_{0}\}$ generated by $\textbf{k}_0$ are explicitly given by:
\begin{equation}
\label{eq:genpw}
\textbf{k}^{i}_{0} = \textbf{k}_0 + n \cdot \textbf{G}_{1} + m \cdot \textbf{G}_{2} + k \cdot \textbf{Q}_{1} + l\cdot\textbf{Q}_{2},
\end{equation}
where different $\textbf{k}_0$'s generate same PW sets with infinite cutoffs due to ergodicity. Here, the index $i$ is in fact $i_{(n,m,k,l)}$, which depends on $n,m, k,l$. For simplicity, here and hereafter, we omit $(n,m,k,l)$ and denote it as $i$. In plotting the coupled PWs, we restrict the integers in Eq.~\eqref{eq:genpw} within a threshold $n_t$, namely $|n, m, k, l| < n_t$. And we adopt a brute cutoff scheme:
\begin{equation}
\label{eq:brute}
|\textbf{k}^i_0|<k_c .
\end{equation}
And generally we have $n_t\cdot|\textbf{G}|= 2\sim 3~k_c$ to achieve satisfying uniformity.

Under small twisted angles, the PWs tend to form clustering sites with a quasi lattice arrangement (not strictly periodic) on the scale of the generating lattice vectors. The size of each cluster is proportional to the threshold $n_t$. In such case, multi-$k$ point sampling is preferred since single $k$ point would require a way too large $n_t$, which results in a extremely large number of PWs to cover the reciprocal space. As the angle increases, the spacing between the points increases and the clusters start to grow in size. With further increase, the sites contact and overlap, resulting in a rather uniform distribution in Fig.~\ref{fig:distr}(b). In this case, single $k$ point sampling would suffice. The above observation also key to draw the quasi-band structure picture.

\begin{figure}[htb!]
\centering
	\includegraphics[width=8.0cm]{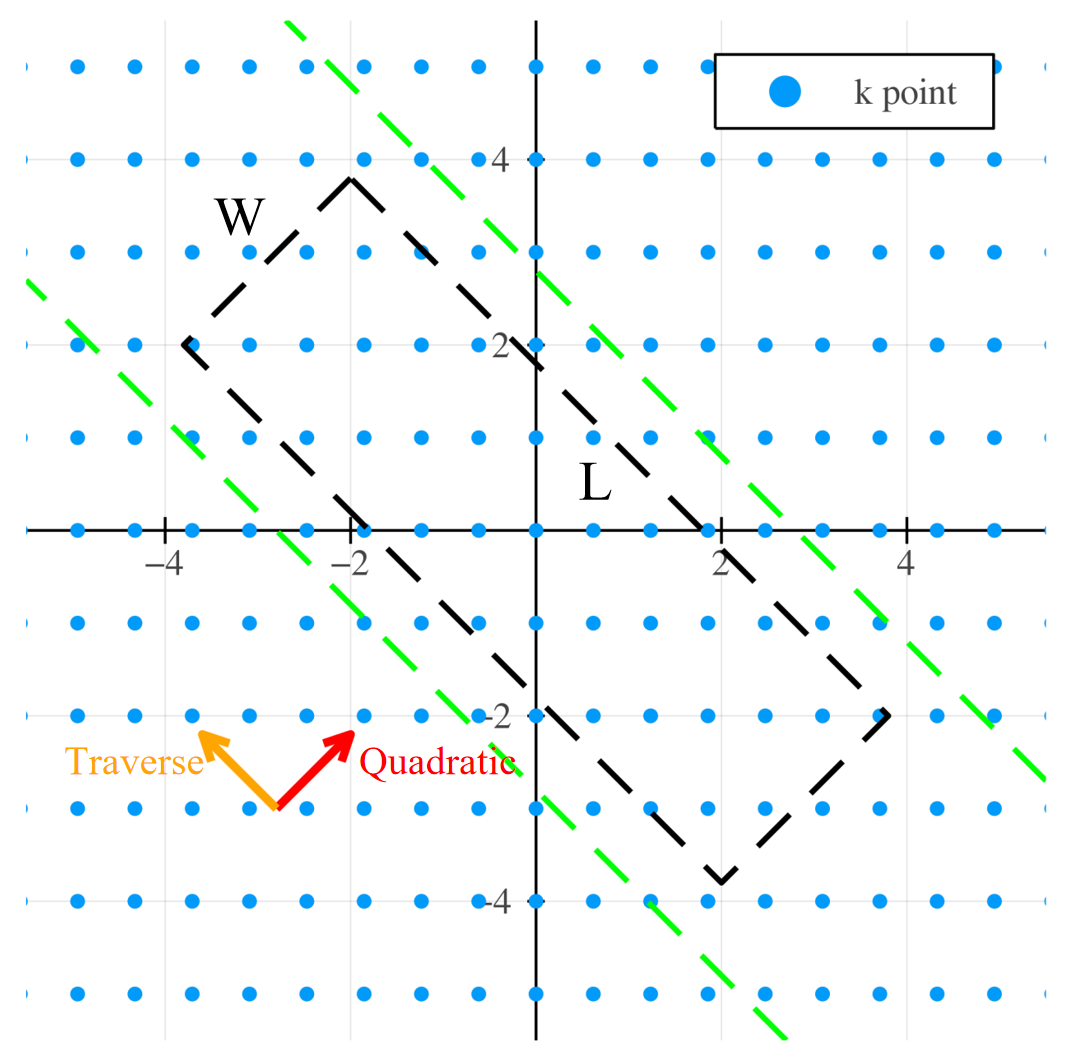}
	\caption{A coupled PW set in 2d reciprocal space of a 2-component 1d incommensurate lattice. The ratio between two lattice constants is $\frac{\sqrt{5}-1}{2}$. The red and orange arrows indicate the quadratic and traverse directions, in which the energies change differently. The green dashed lines and black dashed box represent the brute cutoff scheme and a more effective scheme in \cite{wangtCutoff}, respectively.}
	\label{fig:1d}
\end{figure}

Yet the cutoff scheme can be improved. If we switch to the higher dimensional representation, previous scheme basically introduces a cutoff in some directions of the higher dimensional reciprocal space, while leaving others unaffected. This is the place we can make improvements. Based on our observations in \cite{ChenLoc2021} and more detailed analysis in \cite{wangtCutoff}, the energies of PWs increase quadratically along some directions as in the normal situation, but they also remain almost constant along some other directions. This is better illustrated in a 2-component 1d incommensurate system, as shown in Fig.~\ref{fig:1d}.
Along the diagonal direction, the energies of PWs change quadratically, while along the anti-diagonal direction, they only fluctuate around certain value. Given this nature, one can replace the brute cutoff scheme (green dash) by our newly proposed one (black box), which also truncates from other directions, thus significantly reduce the number of PWs. Detailed analysis and numerics can be found in \cite{wangtCutoff}.

To be more specific for the twisted bilayer system, we define a conjugate $\tilde{\textbf{k}}^i_0$ to each $\textbf{k}^i_0$:
\begin{equation}
\tilde{\textbf{k}}^i_0 = \textbf{k}_0 + n \cdot \textbf{G}_{1} + m \cdot \textbf{G}_{2} - k \cdot \textbf{Q}_{1} - l\cdot\textbf{Q}_{2}.
\end{equation}
Note the sign change compared to Eq.~\eqref{eq:genpw}. New cutoff scheme requires the PWs in $\{\textbf{k}^i_0\}$ and their traverse conjugate satisfy:
\begin{eqnarray}
|\textbf{k}^i_0| &<& k_W ,   \\
|\tilde{\textbf{k}}^i_0| &<& k_L .
\end{eqnarray}
Generally $k_L > k_W$ due to the fact that they converge as $\textbf{e}^{-c k_W}$ and $1/k_L$, respectively \cite{wangtCutoff}. In practical calculations we choose $k_L = 3 \sim 5~k_W$.

\begin{figure}[htb!]
\centering
	\includegraphics[width=15.0cm]{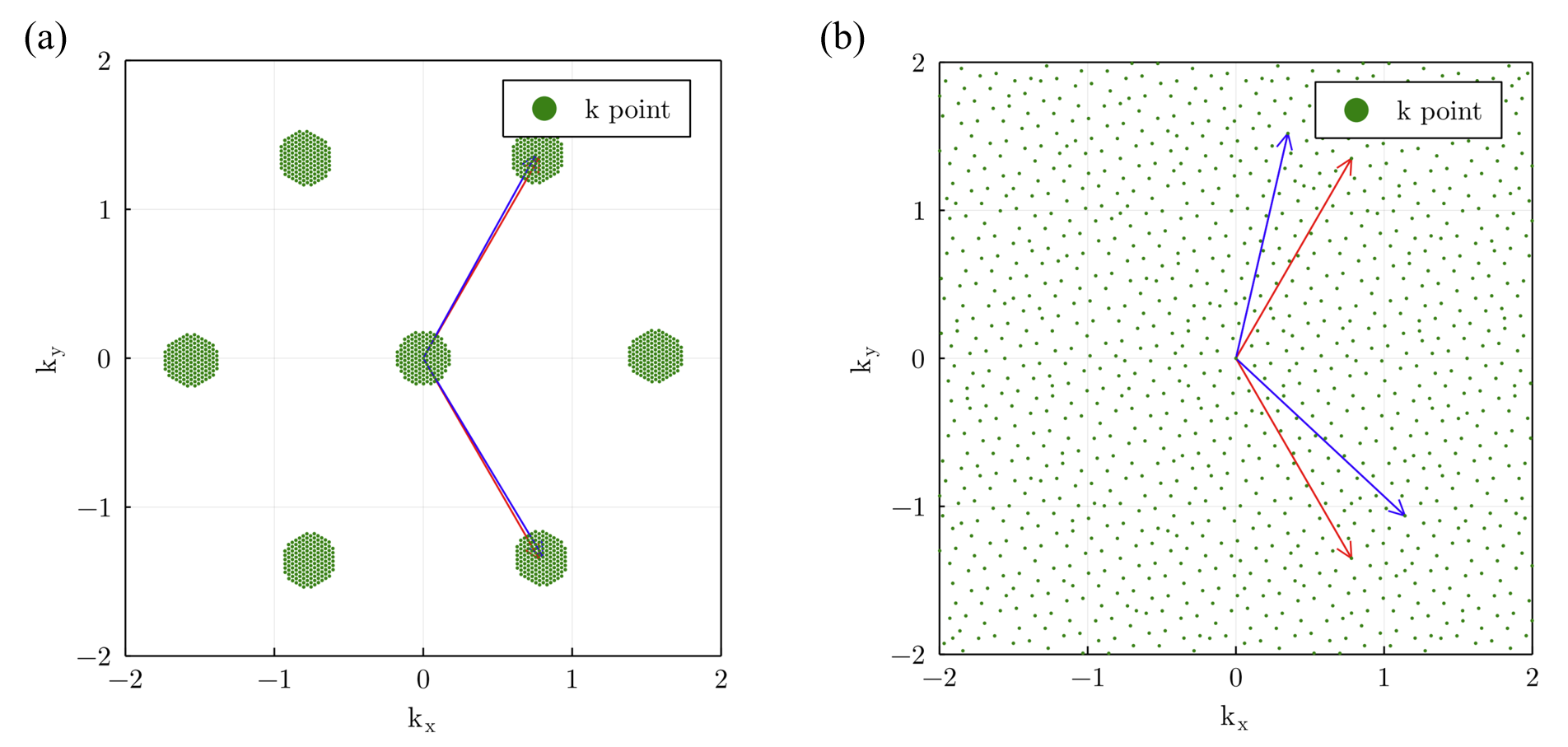}
	\caption{PW distribution in bilayer hexagonal systems with (a) $1.05^\circ$ and (b) $17^\circ$ twisted angles with the new cutoff scheme. In this case, $k_L = 20$ and $k_W = 6$.}
	\label{fig:newdistr}
\end{figure}

The combination of new basis and new cutoffs produces significant reduction in computational cost, making the eigenvalue calculations (either model calculations in this work or the full DFT calculations) within the capability of most modern computers. We illustrate this point by previous example. In Fig.~\ref{fig:newdistr}, we plot the PW distribution with the new cutoff scheme, where $k_L = 20$ and $k_W = 6$.
As shown in the figure, new cutoff scheme has much less PWs. For the $1.05^\circ$ twisted angle, the number of PWs reduces from 42000 to 8000 in this case. To make further estimations, we assume the systems to be bilayer graphene. Together with the localized functions in $z$, the dimension of the Hamiltonian matrix falls in the range of $15000 \sim 80000$ (since for Carbon atom, $2 \sim 10$ localized functions would be suffice). The scale of this matrix is comparable with small to medium systems containing few tens of atoms in the full PW discretization, in drastic contrast with the 11164 atom supercell in the commensurate approximation scheme \cite{Carr2020}.

\section{Quasi-band structures}
Before moving to numerics, we show that under the small twisted angles and weak interlayer coupling limits, a quasi-band structure picture can be formulated, though the ergodicity generally forbids such interpretations and is more fundamental. Here we give a rudimentary analysis under our framework. Though in the following we use TGB lattice as example, the conclusion should be general to other lattice systems.

The quasi-band structure picture largely roots in our previous observation that for small enough twisted angles and not too large $n_t$, the PWs in $\{\textbf{k}^{i}_{0}\}$ distribute in a clustering way, as shown in Figs.~\ref{fig:distr}(a) and \ref{fig:newdistr}(a). If we zoom into each cluster, they further constitutes a hexagonal lattice with the lattice vectors given by:
\begin{equation}
\textbf{m}_{i} = \textbf{G}_i - \textbf{Q}_i \quad i = 1, 2.
\end{equation}
Now we can choose a point in the central unit cell of the cluster (denoted as $\textbf{k}_0$) to generate the coupled PW set. If (1) the PW sets from different generating points are independent, and more importantly, (2) the eigenvalue problem can be well described on these PWs, then we are in a good position to recall the band structure description. Statement (1) is obviously true given a finite of cutoffs. To make statement (2) true, we need the assumption that the interlayer interaction is weak, which generally holds in the layered 2d materials.

\begin{figure}[htb!]
\centering
	\includegraphics[width=15.0cm]{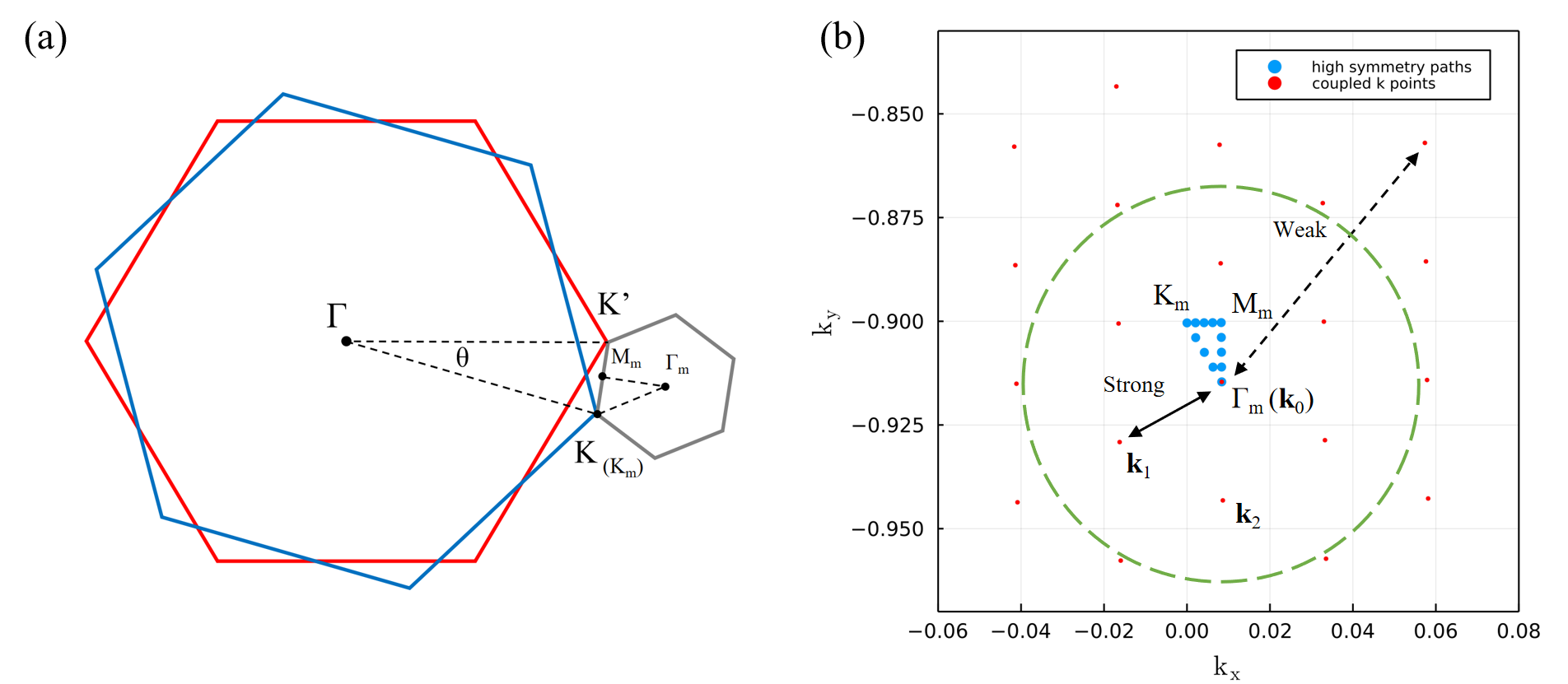}
	\caption{(a) Mini Brillouin zone picture used by continuum model, where the blue and red hexagonals are the BZ of the top and bottom monolayers. (b) Equivalent illustration in our PW framework, where high symmetry paths are defined within the enlarged PW clusters in Fig.~\ref{fig:newdistr}(a). The $\textbf{k}_{0}$, $\textbf{k}_{1}$, $\textbf{k}_{2}$ are the generating k points mentioned in Eq.~\eqref{eq:noint1}. The dashed circle indicates only few neighboring sites are mixed to the eigenstates of a monolayer, due to weak interlayer coupling.}
	\label{fig:mBZ}
\end{figure}

Here we demonstrate it in a heuristic way. For the ease of presentation, we define following notations:
\begin{eqnarray}
\label{eq:noint1}
\{\textbf{k}^{G}_{j}\} &=& \{\textbf{k}_{j} + n \cdot \textbf{G}_{1} + m \cdot \textbf{G}_{2} \} \subset \{\textbf{k}_i \}~, \\
\{\textbf{k}^{Q}_{j}\} &=& \{\textbf{k}_{j} + k \cdot \textbf{Q}_{1} + l \cdot \textbf{Q}_{2} \} \subset \{\textbf{k}_i \}~,
\end{eqnarray}
where each subset contains the PWs only coupled by one of the periodic potentials. By properly regrouping the terms in Eq.~\eqref{eq:genpw}, $\{\textbf{k}^{i}_{0}\}$ can be partitioned using both subsets:
\begin{equation}
\{\textbf{k}^{i}_{0}\} = \{\textbf{k}^{G}_{0}\} \cup \{\textbf{k}^{G}_{1}\} \cup \{\textbf{k}^{G}_{2}\} \cup ... = \{\textbf{k}^{Q}_{0}\} \cup \{\textbf{k}^{Q}_{1}\} \cup \{\textbf{k}^{Q}_{2}\} \cup ... ~ .
\end{equation}
Other generating points, such as $\textbf{k}_1$, $\textbf{k}_2$, are the points in the same cluster of $\textbf{k}_0$, as shown in Fig.~\ref{fig:mBZ}(b). In addition, the full basis is denoted by $\{\textbf{k}^i_0\otimes\xi_u, \textbf{k}^i_0\otimes\xi_d\}$, in which $\{\xi_u\}$ represents the localized basis for the upper layer and $\{\xi_d\}$ the bottom layer.

First assume there is no interaction between the twisted bilayer. Under our basis set, the hamiltonian is block-diagonal in the sense that:
\begin{equation}
\label{eq:hnoint}
H_{full} = \begin{pmatrix} H_{\textbf{k}^{G}_{0}\xi_u} & & & & & \\
                            & H_{\textbf{k}^{G}_{1}\xi_u} & & & &\\
                            & & ... & & &\\
                            & & & H_{\textbf{k}^{Q}_{0}\xi_d} & & \\
                            & & & & H_{\textbf{k}^{Q}_{1}\xi_d} & \\
                            & & & & & ...
 \end{pmatrix} ~,
\end{equation}
where blocks $H_{\{\textbf{k}^{G}_{n}\xi_u\}}$ and $H_{\{\textbf{k}^{Q}_{m}\xi_d\}}$ are the periodic Hamiltonians by Bloch theorem. For now, the upper half block only contains the eigenstates in the top layer and the lower half contains the the eigenstates from the bottom. And these eigenstates can be labeled by the PWs in the central cluster such as $\textbf{k}_0, \textbf{k}_1, \textbf{k}_2$, since there is no off-diagonal term to connect them.

By turning on the interlayer coupling, PWs in the same cluster start to mix. The full hamiltonian matrix now takes the following form:
\begin{equation}
\label{eq:hint}
H_{full}=\begin{pmatrix} H_{\textbf{k}^{G}_{0}\xi_u} & & & & T_{\textbf{k}_{0},\textbf{k}_{1}} & \\
                            & H_{\textbf{k}^{G}_{1}\xi_u} & & T_{\textbf{k}_{1},\textbf{k}_{0}} & &\\
                            & & ... & & &\\
                            & T^{\dagger}_{\textbf{k}_{1},\textbf{k}_{0}} & & H_{\textbf{k}^{Q}_{0}\xi_d} & & \\
                          T^{\dagger}_{\textbf{k}_{0},\textbf{k}_{1}}  & & & & H_{\textbf{k}^{Q}_{1}\xi_d} & \\
                            & & & & & ...
\end{pmatrix} ~.
\end{equation}
In the next section, we will explicitly calculate $T_{\textbf{k}_{i},\textbf{k}_{j}}$. But here we refrain ourselves to qualitative argument. First, the terms in $T_{\textbf{k}_{i},\textbf{k}_{j}}$ are generally small, since they rely on the tunneling between layers, and are $2\sim3$ orders of magnitude smaller compared to diagonal terms (or intralayer hopping) in the 2d materials. Furthermore, to connect nearby $\textbf{k}_{i}$ states in the same cluster, a process with a forward jump of $\textbf{G}_{i}$ to another cluster plus a backward jump of $-\textbf{Q}_i$ to the neighbor of the original site, is needed, which is $O(T^2)$ in scale. One can further imagine the coupling between more distant PWs is even more weak. Therefore, the mixing may only extend to few neighboring sites, as shown in Fig.~\ref{fig:mBZ}(b). And if we cares more about the electronic structures near the central region of the cluster, the new eigenstates can be still well described by the same PW set.

The above analysis underpines the justifiability of quasi-band structure picture, though such picture may only be useful at some specific regions of the original BZ, usually near the Fermi level with some states stand out due to extra symmetries and topological properties. For example, it is of more significance if $\textbf{K}_m$ coincides with the $\textbf{K}$ of monolayer's BZ, otherwise one only gets highly folded band states and a well separated band gap, which is hard to get any further insights. The non-trivial case, which often named as flat bands in the literature, will be discussed in detail next.

\section{Numerical examples}
In this section, we take the bilayer graphene lattice systems as numerical example. For simplicity, we choose the atomic potential to be the one-body, local pseudopotentials. Even within this linear model, the crucial flat band features can be quantitatively reproduced \footnote{We also note that in MacDonald's original paper \cite{MacD2011}, in which the flat band model is first proposed, they used a even more simplified Dirac cone model. We believe linear models better manifest the effect of incommensurate geometries on the emergent non-trivial electronic structures. On the other hand, to study the related correlation phenomena and physics, one definitely has to go beyond single-particle picture.}. Moreover the extension to DFT calculations is straightforward and we reserve the related technical issues, for instance the self-consistent schemes and error analysis, for our future work.

\subsection{Model details}
First, we implement details of potential in Eq.~\eqref{eigen} according to the lattice structure of
graphene. The hexagonal lattice constant is 2.46 \AA, and the interlayer distance is set to 3.52 \AA~\cite{Carr2020}. The atomic potential has chosen to be the combination of the core and local part of the Goedecker-Teter-Hutter's (GTH) pseudopotential \cite{GTHpsp}, which can be written as:
\begin{eqnarray}
\label{eq:rcore}
V_{core}(\textbf{x},z) &=& -b \cdot \frac{Z_{eff}}{r}\textrm{erf}(\frac{r}{\sqrt{2}\xi}), \\
\label{eq:rloc}
V_{loc}(\textbf{x},z) &=& b \cdot \textrm{exp}[-(r/\xi)^2]\times[C_1 + C_2\cdot(r/\xi)^2],
\end{eqnarray}
where $Z_{eff}, \xi, C_1, C_2$ are fitted parameters and $r = \sqrt{|\textbf{x}|^2 + z^2}$. The nonlocal part has also been neglected for simplicity. The Fourier transform of the above potentials can be calculated analytically:
\begin{eqnarray}
\label{eq:kcore}
V_{core}(\textbf{K}) &=& -b \cdot 4\pi\frac{Z_{eff}}{\Omega}\frac{\textrm{e}^{-(K\xi)^2/2}}{K^2}, \\
\label{eq:kloc}
V_{loc}(\textbf{K}) &=& b \cdot \sqrt{(2\pi)^3}\frac{\xi^3}{\Omega}\textrm{e}^{-(K\xi)^2/2}\{C_1 + C_2\cdot[3-(K\xi)^2]\},
\end{eqnarray}
where $\Omega$ is the volume in which PWs are normalized. In the following we are going to use a limited number of basis and to circumvent that, a scaling factor $b$ is used to reproduce the band structures of the monolayer graphene.

Since the $2p_z$ orbitals are the most involved atomic states in the flat bands near the Fermi level, we choose a minimum of 2 localized functions in the $z$ direction, one for the upper layer and one for the bottom. Based on symmetry, these two should also have the same form, denoting as $\xi_{u}(z)$ and $\xi_{d}(z)$ respectively. Due to the weak interlayer coupling, for the crossing terms between the states in different layers, only $\bra{\xi_{i}} V_{i} \ket{\xi_{j}}$ give non-zero contribution, with $V_{i}$ labeling the potential in the corresponding layer and $j$ labeling the other layer. The kinetic terms between different layers and the potential terms like $\bra{\xi_{i}} V_{j} \ket{\xi_{i}}$ are taken as zero. Further, we can neglect the integrals in the square bracket of Eq.~\eqref{kin} from the same layer, since this is only a constant shift in the eigenvalues.

Now we are ready to numerically evaluate the Hamiltonian matrix. We will proceed in the following two ways. And after sketching them, results will be shown and discussed.

\subsubsection{Parameterization treatment}
A simple way to treat the integrals in Eqs.~\eqref{kin} and \eqref{prec} is to parameterize them according to band structure of monolayer graphene and the interlayer coupling strength. This treatment has the advantage that there is no need to actually construct localized functions and evaluate the integrals, but the extendability of the model is somehow weakened.

First, for the monolayer, the integrals in Eq.~\eqref{preal} can be approximated by:
\begin{equation}
\label{eq:mono}
\<\Omega_{j0}|V_{mono}|\Omega_{i0}\> = \frac{\delta_{\textbf{G}_{lm},\textbf{k}_{ij}}}{A_c}S(\textbf{k}_{ij}) \int^{\infty}_{-\infty} V^{at}(\textbf{k}_{ij},z) \xi^*_0(z) \xi_0(z) \dd z \approx b\cdot\delta_{\textbf{G}_{lm},\textbf{k}_{ij}}S(\textbf{k}_{ij}) V_{core+loc}(\textbf{k}_{ij})~,
\end{equation}
where $\textbf{G}_{lm} = l\cdot \textbf{G}_1 + m\cdot \textbf{G}_2$ is the compact notation of monolayer's reciprocal lattices. And we have reproduced the band structure\footnote{Specifically, we reproduce the absolute position of the Dirac point as well as the band width of $p_z$ band.} of monolayer graphene by setting $b = 0.0097$.

With that, in the twisted bilayer case, the matrix elements of potential can be derived as:
\begin{eqnarray}
\<\Omega_{jm}|V|\Omega_{in}\> & = & \sum_{rs}\frac{\delta_{\textbf{G}_{rs},\textbf{k}_{ij}}}{A_c}S_u(\textbf{k}_{ij}) \int^{\infty}_{-\infty} V^{at}_{u}(\textbf{k}_{ij},z) \xi^*_m(z) \xi_n(z) \dd z \\
 & & +~ \sum_{uv}\frac{\delta_{\textbf{Q}_{uv},\textbf{k}_{ij}}}{A_c}S_d(\textbf{k}_{ij}) \int^{\infty}_{-\infty} V^{at}_{d}(\textbf{k}_{ij},z) \xi^*_m(z) \xi_n(z) \dd z \\
& \approx & b \cdot \frac{\delta_{mu}\delta_{nu}\delta_{\textbf{G}_{rs},\textbf{k}_{ij}}S_u(\textbf{k}_{ij}) + \delta_{md}\delta_{nd}\delta_{\textbf{Q}_{uv},\textbf{k}_{ij}}S_d(\textbf{k}_{ij})}{A_c} V_{core+loc}(\textbf{k}_{ij})\\
\label{eq:param}
& &  + ~ t \cdot b \cdot \frac{\delta_{mu}\delta_{nd}\delta_{\textbf{G}_{rs},\textbf{k}_{ij}}S_u(\textbf{k}_{ij}) + \delta_{md}\delta_{nu}\delta_{\textbf{Q}_{uv},\textbf{k}_{ij}}S_d(\textbf{k}_{ij})}{A_c} V_{core+loc}(\textbf{k}_{ij}) ~ ,
\end{eqnarray}
where in the last two lines we have regrouped the terms such that in the 3rd line we have the intralayer coupling terms and 4th line the interlayer coupling terms. Previous integrals have also been replaced the integrals by $b$ in Eq.~\eqref{eq:mono} and a interlayer tunneling parameter $t$. In this treatment, we take $t = 0.01$ which best reproduces most features of the flat bands.

This choice can be compared with the paremeters in the TB model, where the intralayer hopping strength is $\sim 3.0$ eV and interlayer tunneling is $\sim 0.1$ eV \cite{Andrei2020,MacD2011,Mac13GraphParam}. Notice that the hopping strength in TB contains the contributions both from kinetic and potential operators. To estimate the corresponding $t'$ in TB model, we make use of the Viral theorem and roughly assume the kinetic contribution is approximately $-\frac{1}{2}$ of the potential contribution in the hopping term. Therefore $t' \approx 0.1/6 = 0.0167$, in good agreement with our choice of $t = 0.01$.

\subsubsection{Atomic-like orbitals}
Alternatively, we can explicitly choose localized part in the form of $n=2$ shell atomic radial functions, which can be written in the atomic units as \cite{zengQuantum}:
\begin{equation}
\label{eq:xiz}
\xi_{at}(z) = C \cdot z \cdot \textrm{exp}[-3 z]~,
\end{equation}
where $C$ is the normalized factor. Specifically we use Eq.~\eqref{prec} to calculate the matrix elements since the analytical form of the localized functions is available. In the monolayer calculations, $\xi(z)$ does not come into the kinetic terms if we drop the integral part in Eq.~\eqref{kin}, which is only a constant shift to eigenvalues. The potential matrix elements only require the integrals between the localized functions of the same layer, and take the form:
\begin{equation}
\label{eq:pintra}
\<\Omega_{j0}|V_l|\Omega_{i0}\> \sim \int^{\infty}_{-\infty}  \frac{V_l(\textbf{k}_{ij},k_z)}{(3 + \textrm{i}\cdot k_z)^3} \frac{\dd k_z}{2\pi} ~,
\end{equation}
where $V_l$ is given by Eqs.~\eqref{eq:kcore} and \eqref{eq:kloc}. We have reproduced the band structures by setting $b_{l} = 10.277$, where a subscript $l$ is used to distinguish from $b$ in the previous treatment. The normalized factor $C$ has also been absorbed in it.

For the twisted bilayer, extra potential terms from interlayer overlap take the form:
\begin{equation}
\label{eq:pinter}
\<\Omega_{ju}|V_{u, d}|\Omega_{id}\> \sim \int V_{u, d}(\textbf{k}_{ij},k_z) \cdot\textrm{exp}(-1.5d_z - \textrm{i}k_z d_z) \cdot \frac{-2\textrm{i}+k_z d_z+\textrm{e}^{\textrm{i}k_z d_z}(2\textrm{i}+k_z d_z)}{k_z^3} \dd k_z~,
\end{equation}
where $d_z = 3.52~\AA$ is the interlayer distance as defined previously. In the above, we only include the
contributions from the interlayer region due to a fast decay of the localized function from its center. Terms like $\<\Omega_{ju}|V_{d}|\Omega_{iu}\>$ or $\<\Omega_{jd}|V_{u}|\Omega_{id}\>$ are also dropped for similar reason.

\subsection{Results and discussions}
With previous efforts, all calculations can now be performed on PC with 16G RAM, using Julia as programming language \cite{Julia2017}. The eigenvalue problems are solved by calling the spectral decomposition of LinearAlgebra packages. Moreover, we use adaptive Gauss-Kronrod quadrature methods for the numerical evaluations of Eqs.~\eqref{eq:pintra} and \eqref{eq:pinter} \cite{gkmethod}.

\begin{figure}[htb!]
\centering
	\includegraphics[width=15.0cm]{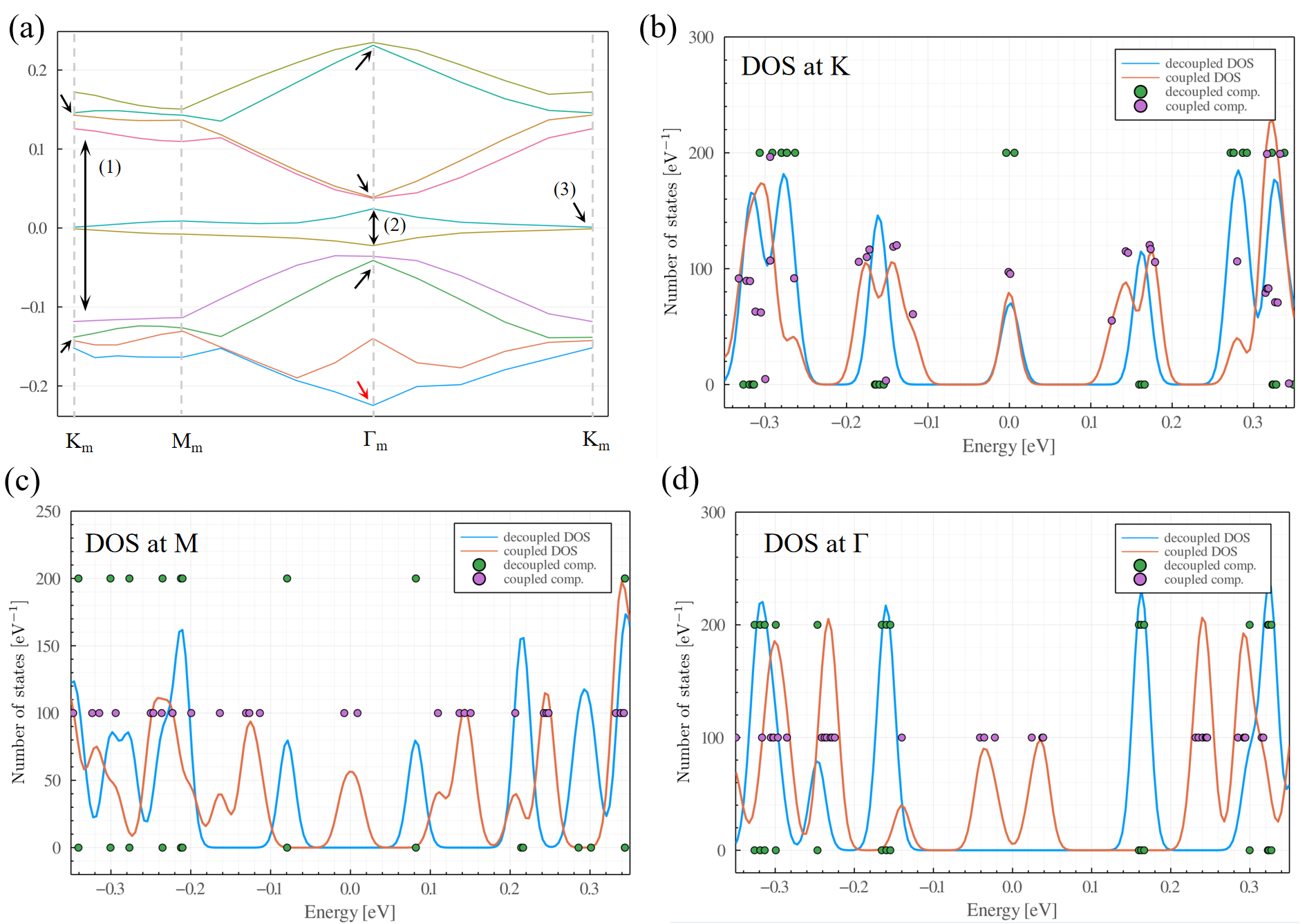}
	\caption{The calculation results from parameterization treatment. (a) Band structures along high symmetry paths of the mini
BZ. Numbers and arrows indicate some important features of these bands. DOS plots at special points: (b) $\textrm{K}_m$, (c) $\textrm{M}_m$ and (d) $\Gamma_m$.}
	\label{fig:bsdosi}
\end{figure}

\begin{figure}[htb!]
\centering
	\includegraphics[width=15.0cm]{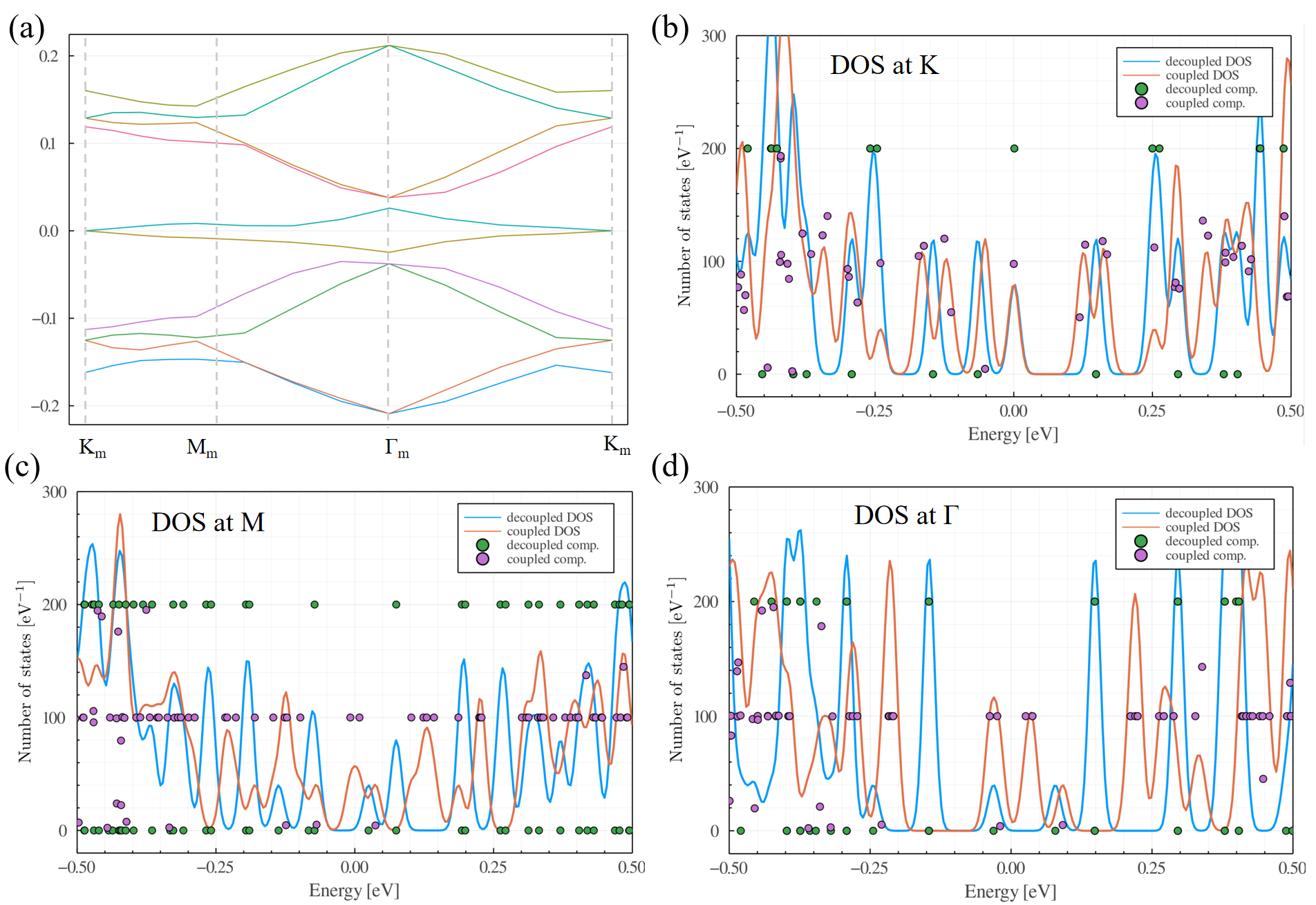}
	\caption{The calculation results from atomic-like orbitals. (a) Band structures along high symmetry paths of the mini
BZ. DOS plots at special points: (b) $\textrm{K}_m$, (c) $\textrm{M}_m$ and (d) $\Gamma_m$.}
	\label{fig:bsdosl}
\end{figure}

The plots of band structures and density of states (DOS) from both treatments are shown in Figs.~\ref{fig:bsdosi} and \ref{fig:bsdosl}, respectively. Ten bands near the Fermi level have been shown. To better illustrate the effect of interlayer coupling, DOS's with (purple circles in the middle) and without (green circles in the upper or below) interlayer couplings are shown. The upper and lower circles represent the states from in the top and bottom layers respectively. While the middle circle's position quantitatively reflects the relative contribution from each layer: if it is higher from the exact middle, then it has more contribution from the top-layer states and vice versa.

Here we focus on the flat bands. A closer look at them shows that it is the interlayer couplings that mixes top and bottom layer states which are close in energy. Most states have both contributions from the top and bottom layer, but one can see that these flat bands are well separated from the rest, highly folded states. From both treatments, the calculated band structures have many features in good agreements with other theoretical results and experiments, including: (1) at $\textrm{K}_m$, a separation of $\sim$ 0.2 eV between the flat bands (here a two-fold degenerate states) and other bands; (2) at $\Gamma_m$, the flat bands disperse with a width of 50 meV, falling into range of $30 \sim 80$ meV from experiments and other calculations \cite{Andrei2020,Carr2020,Li2010,massat2021}; (3) Degeneracies and trends are generally consistent with the DFT results in Fig.~5(a) of \cite{Carr2020}, and the particle-hole symmetry is also respected near the Fermi level. These are indicated by the arrows and corresponding numbers in Fig.~\ref{fig:bsdosi}(a), and also applies to Fig.~\ref{fig:bsdosl}(a).

\begin{figure}[htb!]
\centering
	\includegraphics[width=15.0cm]{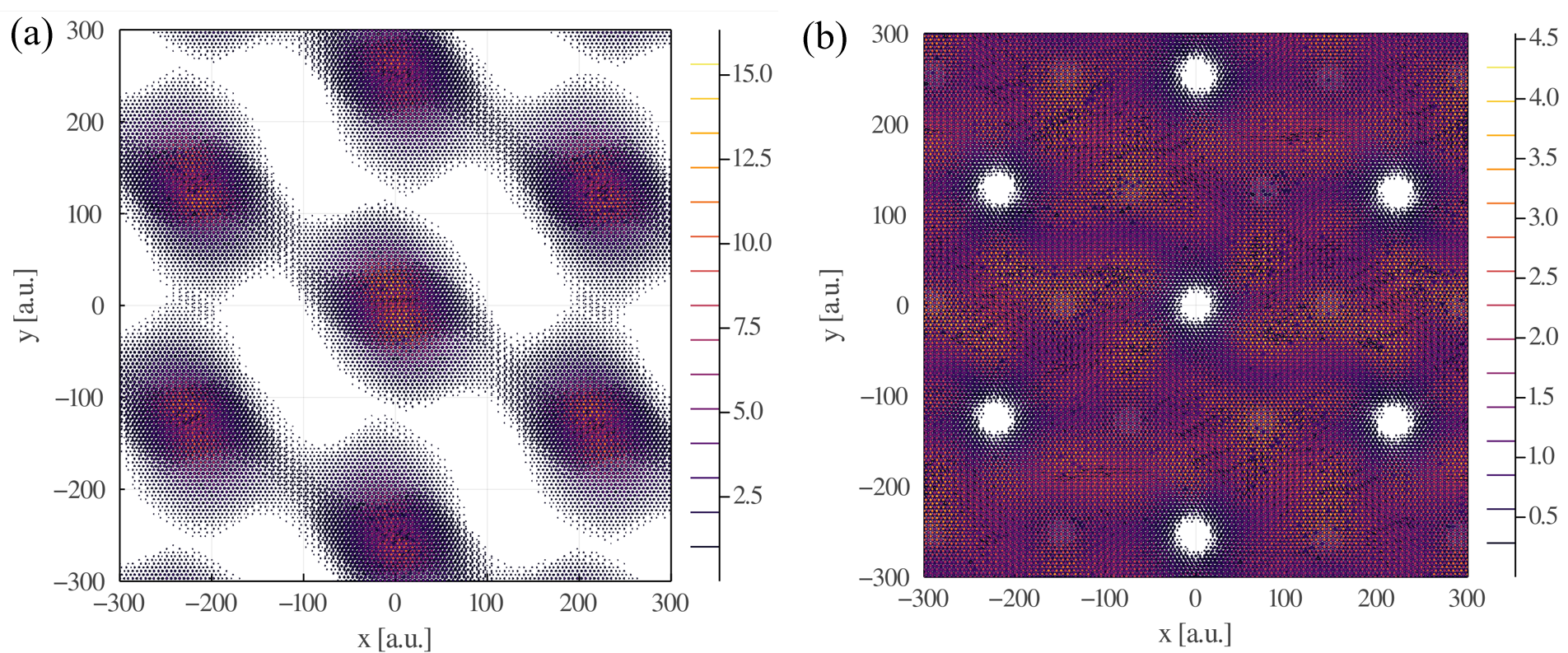}
	\caption{The density distribution of the highest occupied states at (a) $\textrm{K}_m$ and (b) $\Gamma_{m}$.}
	\label{fig:dens}
\end{figure}

The density distributions in the top layer of the highest occupied states at $\textrm{K}_{m}$ and $\Gamma_{m}$ are plotted in Fig.~\ref{fig:dens}. These distributions are highly renormalized on the scale of $1/|\Delta \textbf{k}|$ (or the Mori\'{e} length scale), where $|\Delta \textbf{k}|$ is the spacing between the coupled PWs. Interestingly, the stripe distribution in Fig.~\ref{fig:dens}(a) has also been observed in the scanning tunnelling spectroscopy measurements, where the local net charge is found to have similar pattern on the same length scale \cite{Jiang2019}. Even though such comparison is not strict, the density plots unambiguously show the precursors of strongly correlated physics.

Now we comment on the efficiency and extendability. Firstly, the computational cost has been significantly reduced. For each $k$ point calculations, it takes about 10 minutes to finish and the most time-consuming part is the diagonalization of the matrix. And we can further compare with the commensurate cell scheme, which requires 11,164 atoms supercell and consumes 3,100,000 core hours for the magic angle TBG \cite{Andrei2020}. The fact that DFT and self-consistent field calculations are more complicated does not alter our conclusion. Needless to say we also have plenty room to improve the efficiency, for instance using GPU hardwares, implementing parallelizations and effective iterative eigensolvers. Then we discuss the extendability. Currently we use fitted parameters to counteract the fact that we are not using large enough localized basis and to simplify some calculations. However, there are at most two parameters, and without too much efforts, such calculations can be made parameter-free. Also note that recently there have risen novel electronic methods including the Fermionic neural networks Ansatz \cite{ferminet20} and automatic differentiation machinery \cite{ad22}, which would help to make our calculations more effectively \emph{ab initio}. Therefore, we believe our framework is more extendable compared to the traditional model hamiltonians and tight binding models.

Below are some further remarks. Even though the results from two treatments look identical, there are still minor places in which they differ. As shown by the red arrow in Fig.~\ref{fig:bsdosi}(a), parameterization treatment fails to predict a degenerate state here, and a closer look also indicates that the degeneracies at other points are not strictly obeyed. In contrast, using the atomic-like orbitals better preserved such symmetries. This can be understood by the fact that the parameterizations in Eq.~\eqref{eq:param} oversimplifies the the integrals in $z$, and might break some $z$ related symmetries in the process. While in the atomic orbital treatment, we explicitly calculate the related integrals and have also pre-assumed some symmetries in the form of the localized basis, which help to better preserve the symmetries in the band structures.

Last, we are zooming into 10 bands in the vicinity of the Fermi level out of thousands of bands to provide previous band structure and DOS plots. Also in the reciprocal space we are only sampling regions near the $K$ point. Therefore even though the flat bands are extremely important to the low energy physics of the system and stand out by some symmetry-protected topology \cite{TGBAnaly1,TGBAnaly2}, they are only miniscule part of the total electronic structure. And the implication is that model Hamiltonians that only reproduce the flat bands plus nearby bands cannot appropriately descrbile the properties of the whole system. Yet in our framework we have provided a systematic way to fully sample the reciprocal space and study the total electronic structure.

\section{Conclusion}
In this paper, we focus on the practical calculations of twisted 2d material systems and introduce extensions of our PW framework in the following aspects: (1) a tensor-producted basis set with PWs in the incommensurate dimensions and localized functions in $z$ direction, (2) the practical application of our newly developed cutoff techniques, and (3) a quasi-band structure picture under the small twisted angles and weak interlayer coupling limits. With (1) and (2), we have remarkably reduced the dimensions of hamiltonian matrix, which makes the electronic structure calculations of twisted bilayer 2D material systems affordable to most modern computers. And (3) helps us better organize the calculations as well as understand results. We further use the linear TGB system with magic twisted angles ($\sim 1.05^{\circ}$) as numerical examples. We have reproduced the famous flat bands with key features in good quantitative with other theoretical and experimental results. In terms of efficiency, our framework has much less computational cost compared to the commensurate cell approximations. While it is also more extendable compared to the traditional model hamiltonians and tight binding calculations. Lastly, nonlinear terms like Hartree energy and exchange-correlation energy can be readily included in the framework thus more effective and accurate DFT calculations of incommensurate 2D material systems can be expected in the near future.

\section{Acknowledgement}
The authors would like to thank the valuable discussions with Prof.~Huajie Chen, Dr.~Ting Wang and Prof.~Daniel Massatt. This work was partially supported by National Key R {\&} D Program of China under grants 2019YFA0709600 and 2019YFA0709601. Y. Zhou's work was also partially supported by the National Natural Science Foundation of China under grant 12004047.

\section*{References}
\bibliography{zhou}

\end{document}